\documentclass[sigconf,authorversion]{acmart}
\usepackage{tabularx}
\usepackage{balance}

\AtBeginDocument{%
  \providecommand\BibTeX{{%
    \normalfont B\kern-0.5em{\scshape i\kern-0.25em b}\kern-0.8em\TeX}}}

\author{Jiachen Jiang}
\affiliation{%
  \institution{Dartmouth College}
  \city{Hanover}
  \state{New Hampshire}
  \country{USA}}
\email{jcjiang42@gmail.com}

\author{Soroush Vosoughi}
\affiliation{%
  \institution{Dartmouth College}
  \city{Hanover}
  \state{New Hampshire}
  \country{USA}}
\email{soroush.vosoughi@dartmouth.edu}


\copyrightyear{2020}
\acmYear{2020}
\setcopyright{acmlicensed}\acmConference[FATE/MM'20]{2nd International Workshop on Fairness, Accountability, Transparency and Ethics in MultiMedia}{October 12, 2020}{Seattle, WA, USA}
\acmBooktitle{2nd International Workshop on Fairness, Accountability, Transparency and Ethics in MultiMedia (FATE/MM'20), October 12, 2020, Seattle, WA, USA}
\acmPrice{15.00}
\acmDOI{10.1145/3422841.3423534}
\acmISBN{978-1-4503-8148-2/20/10}
\begin{document}
\fancyhead{} 


\title{Not Judging a User by Their Cover} 
\subtitle{Understanding Harm in Multi-Modal Processing within Social Media Research}


\begin{abstract}
  Social media has shaken the foundations of our society, unlikely as it may seem. Many of the popular tools used to moderate harmful digital content, however, have received widespread criticism from both the academic community and the public sphere for middling performance and lack of accountability. Though social media research is thought to center primarily on natural language processing, we demonstrate the need for the community to understand multimedia processing and its unique ethical considerations. Specifically, we identify statistical differences in the performance of Amazon Turk (MTurk) annotators when different modalities of information are provided and discuss the patterns of harm that arise from crowd-sourced human demographic prediction. Finally, we discuss the consequences of those biases through auditing the performance of a toxicity detector called Perspective API on the language of Twitter users across a variety of demographic categories. 
\end{abstract}

\maketitle

\section{Introduction}

The Information Age has generated unprecedented levels of human innovation and discovery, but it has also led to the rise of a startling array of digital injustices. In particular, the rise of harmful content on social media has affected everything from mass shootings to national elections. Furthermore, the anonymity of online communication has challenged existing legal frameworks that regulate hate speech and harmful language. 

Recent controversy has arose surrounding how various social media companies have decided to reckon with harmful content hosted on their platforms. Conversation has focused on the moderation systems of Facebook, Twitter, Reddit, and others that automate toxicity detection and removal for the massive amount of content posted to their servers daily. Traditional toxicity detectors check new content against existing databases of profane or harmful text. One issue with this approach is that a majority of terms associated with hate speech are themselves identifiers for marginalized identities \cite{waseem16}. For example, homophobia and racism present in much of toxicity data sets can mean that words such as "lesbian" or "black" will become classified as features of hate speech. When automatic hate speech systems remove content that contains these words, they disproportionately remove content by and about members of those same marginalized communities who use the terminology to self-identify or share their lived experiences. 

Though social media research is associated mainly with natural language processing, we discuss in this paper why multimedia processing and its ethical considerations are of interest to social media researchers. Experiment design for crowd-sourcing data annotation is often an afterthought in research. We identify statistical differences in the performances of MTurk annotators when different modalities of information are provided and discuss patterns of harm that arise from crowd-sourcing human demographic prediction. We then discuss the consequences of those biases by walking through our audit of Perspective API, a toxicity classification model developed in collaboration by research teams in Jigsaw and Google. We determine statistically significant differences between the API's performance on the language of Twitter users across a variety of demographic categories. Finally, we talk through the significance of these results for the field of social media research and describe next steps in the project. Our research seeks not to offer concrete if arbitrary definitions of right and wrong, rather to question long-held assumptions surrounding harm and the elimination thereof within computational systems. We believe the lessons detailed in this paper are important to note not just for researchers working with multimedia or natural language processing, but for the computer science community as a whole.

\subsection{Machine Learning in Social Media Moderation}

Modern hate speech detection systems used by Facebook, Twitter, Reddit, and others appear to be - at least initially - far more sophisticated than the straightforward processes described above. They employ complex machine learning techniques to classify content as hate speech or otherwise harmful. These toxicity detectors have run into general obstacles of natural language processing, however, such as human biases within word embeddings \cite{bolukbasi16} and imperfect training data that do not paint an accurate picture of the world. Even more concerning is the fact that these algorithmic systems are severely limited by their lack of interpretability. Due to the massive amount of training data and the black-box nature of complex computational processes, it is difficult if not impossible to explain the rationale behind the decisions that these systems make. In other words, the systems built to automatically censor a person's speech often do not have room for a manual appeal process that can hold them accountable. Perhaps most difficult of all to address is the lack of clear intent and impact, which means that existing legal definitions and frameworks surrounding discrimination and harm will falter when computational tools are the perpetrators of injustice. 

What are the consequences of toxicity detection gone wrong? Research has shown that tweets written in African-American English are twice as likely to be labeled as offensive by models trained on popular hate speech data sets than those in Standard American English, which not only perpetuates African-American stereotypes about aggression and vulgarity, but actively excludes voices of a marginalized community from research and study \cite{sap19}. Another fallout from the lack of transparency and accountability in automated hate speech detection systems can be seen in a 2019 lawsuit against YouTube for restricting all LGBT+ content to mature audiences \cite{bensinger19}.

The vast majority of repercussions, however, are felt by users without much platform at all. They use their personal social media to levy complaints against Facebook, among other forums, where users have been given suspensions and had their posts removed for discussing their lived experiences of discrimination. In everything from Medium articles to Twitter threads, people of color describe the many measures they take to avoid being suspended on social media, from back-up accounts to retain access to important groups or pages, to a buddy system to inform the broader community when a user has been silenced. Despite the prevalence of these complaints on social media, there exists little to no academic literature investigating the issue \cite{guynn_2019}. In this way, automatic hate speech detection systems appear only a few irresponsible deployments away from becoming an incredibly effective vehicle of censorship against entire communities - all while operating under the guise of maintaining civility. 

\section{Human Demographic Identification} 

Though social media moderation may seem initially to be a problem  processing of natural language, it relies heavily on work conducted in the field of multimedia processing.

Social media data is the primary subject of most modern NLP research due to its abundance and accessibility; human subject protection laws and frameworks are still a long way away from fully regulating the intricacies of social media research. It has been used for everything from predicting the outcomes of political elections \cite{burnap_gibson_sloan_southern_williams_2015} to crime prediction  \cite{sloan_morgan_2015}, to wildly varying degrees of success. By its nature, however, social media research works almost entirely with public data that contains little to no concrete identifiers of identity. Even when those indicators exist, the opt-in nature of sharing information on social media can result in biased data sets. For example, the users who share information about geographic location are significantly different from the overall population of users [29]. 

As a result, demographic information must be obtained for hundreds of thousands (and more) of individuals for even a single trial. Their source usually comes down to either human annotation or prediction by a computational system. The latter option is magnitudes cheaper and faster, leading to an explosion of research describing how to predict the demographics of Twitter users using everything from regression models built on website traffic data \cite{culotta_ravi_cutler_2016} to text analysis of usernames \cite{wood-doughty_andrews_marvin_dredze_2018} to recursive neural networks \cite{kim_xu_qu_wan_paris_2017}. The performance of these models are ultimately validated on a smaller set of Twitter users labeled individually for ethnicity and gender by human annotators, either the researchers themselves or - far more commonly - tens of thousands of anonymous strangers on crowd-sourcing platforms.

Despite criticism and controversy regarding human subject privacy and inaccuracies in prediction, it is clear that computational demographic prediction is and will continue to be an integral part of natural language processing and broader computer science research. The general consensus in the community seems to be that despite inaccuracies and bias, the ability to predict demographic information is too crucial to research to give up entirely. Not only are user demographics necessary to make conclusions applicable to the real world, they must be known for disparate impact to be identified and eliminated. Data without any features relating to personal identity is the computational equivalent of a colorblind society - marginalization and censorship continue to occur due to discrimination by proxy, they just become invisible and unassailable. At the heart of human demographic prediction is a difficult choice between attempting to diagnose the problem using a problematic process or allowing disparate impact to remain unseen. 

\subsection{Amazon Mechanical Turk}

Due to the time and financial limitations of our research, we settled on using Amazon MTurk for our data annotation. It was important to us that despite the ethical and accuracy concerns that came with that choice, however, that we base our experiment design on data as opposed to our personal assumptions. Though there is no such thing as a perfect experiment design, especially when working with such thorny topics of bias and fairness, we wanted to be able to justify our choices. In particular, we investigated how the type and amount of Twitter user account information presented to workers on crowd-sourcing platforms affect their resulting annotations.  

Most prompts for human demographic prediction simply provide all the information available to researchers, whatever that is - usually a combination of images and text descriptions. In some cases, a small collection of the user's tweets are also included. Data annotators then make decisions based on multimedia information. Even when researchers provide just a single type of media such as text on its own, they rarely discuss or take into account additional considerations that stem from making that choice. For example, annotators asked to make a decision on demographics using only visual media may overemphasize the user's appearance and self-presentation. On the other hand, annotators working off of only text information may describe the user's sociolinguistic identity instead of the real demographics of the user. There is no doubt that the harms and disparate impact that arise from multimedia processing are central to any and all research that includes human demographic identification, which includes much of social media research.

\subsection{Methods} 

We conducted A/B testing using collections of a hundred Twitter accounts scraped from the TweePy Stream in real-time. As we cannot know the real demographics of these users, we focused our analysis on answering one key question: are the annotations of one category statistically different from those of another? What patterns of disparate impact arise when humans make a decision based on different modalities of data? Our three trials were: 

\begin{enumerate}
    \item Full Twitter profile, minus the exact username so to protect the privacy of the users and to discourage participants from accessing additional information for demographic prediction and biasing our results. 
    \item Only the text portions of the Twitter profile, which included the display name and description.
    \item Only the image portions of the Twitter profile, which included the cover photo and profile picture. 
\end{enumerate}

We ran three trials, each with the same 100 Twitter profiles with differing types of information. Five participants annotated each user, each for a total of 500 annotations.

\subsection{Results}

We grouped our annotations by the demographic categories of "Age," "Race," and "Gender." Though we primarily used the grouping discussed in greater detail during our audit of Perspective API below, we simplified the Race demographic to simply "White" and "Non White" so to incorporate the broad range of personal definitions of race that annotators likely have. We then calculated the \textit{entropy} for each account annotated to determine the level of agreement across the five separate MTurk annotators, such that a lower value of entropy denotes more agreement between the annotators. Finally, we conducted two-sided T-tests between each pair of categories across demographics. The results can be seen in Table \ref{pvalues}.

\begin{table}[h]
\large
\begin{tabular}{|l|c|c|c|l}
\cline{1-4}
                       & \textbf{Age} & \textbf{Race} & \textbf{Gender} &  \\ \cline{1-4}
\textbf{All vs Image}  & 0.58       & 0.71        & 0.91          &  \\ \cline{1-4}
\textbf{All vs Text}   & 0.69       & 0.72        & 0.03*         &  \\ \cline{1-4}
\textbf{Text vs Image} & 0.87       & 0.99        & 0.01*         &  \\ \cline{1-4}
\end{tabular}
\caption{$p$-values of two-sided t-tests across age, race, and gender. Values less than 0.05 are statistically significant. * < 0.05 }
 \label{pvalues}
\end{table}

Though differences exist between modalities for Age and Race, the most interesting results appear for the Gender demographic. Though the annotations done by MTurk workers with access to both image and text information aligned very closely to those with access to only image (\textasciitilde 0.9085), there was great disparity between those categories and the annotations conducted by workers with only access to text, with p-values well under the threshold for statistical significance. The mean entropy of gender annotations across modalities can be seen in Figure \ref{barplot} below.

 \begin{figure}[h]

 \includegraphics[width=0.99\columnwidth]{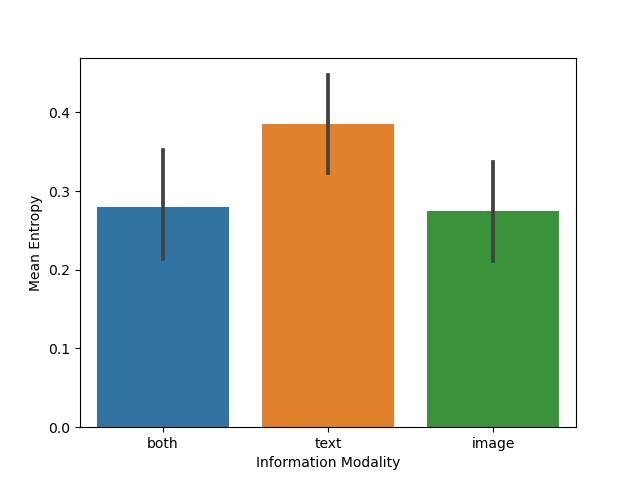}
 \caption{Mean entropy of gender annotations across modalities, with standard error. A lower value of entropy denotes greater agreement across annotators.}
 \label{barplot}
 \end{figure}

Gender is a social construct that depends hugely on presentation and appearance, meaning that a user's text description may vary greatly from their provided images. We see that the personal understanding of gender that MTurk workers annotate based off of varies greatly in the context of visual versus textual information. In addition, we note that image appears to be more polarizing of an information medium than next. There is little difference between the annotations of MTurk workers with access to only images as opposed to both image and text, suggesting that when they have access to both modalities of information, they still base their judgement much more on the provided image. That is, assumptions and information gained from visual data overshadow those that are gained from natural language data.

\subsection{Next Steps}

Though these results are not definitive and should not be taken as proof of the phenomenon, they suggest a need for a more critical understanding of how experiment design choices in regards to crowd-sourced data annotation can drive harm and general biases within the data. Our immediate next step is to identify the most ambiguous and controversial accounts, so to fully understand who is harmed by these system behaviors and in what ways. 

Social media research is a field built on crowd-sourced human demographic prediction. If there exist systematic biases in the processes through which much of our data is obtained, then there must also be systematic biases in the tools we build and the conclusions we reach.

\section{Auditing Perspective API}

We discuss the consequences of the behavior above through the Perspective API, a free tool developed by Jigsaw and Google’s Counter Abuse Technology team as part of a collaborative research project. Its primary and oldest offering is a machine learning model trained on hundreds of thousands of pieces of human-annotated text \footnote{https://www.perspectiveapi.com/home} to predict the perceived harm of a comment on its readers. 

\subsection{Background}

At its conception, Perspective operated very similarly to other toxicity detection tools, ”[providing] a score from zero to 100 on how similar the new comments are to the ones identified as toxic.” Currently, they offer a variety of additional tools ranging from real-time feedback for commenters to grouping comments by topics for readers. 

Though at the moment Perspective only offers production models for ’Toxicity’ and ’Severe Toxicity’, experimental models like ’Flirtation’ and ’Identity Attacks’ are  available and suggest that future versions may contend with problems of sexual assault and hate speech - high-profile issues not only for their specificity but for their strong precedence in the legal field. Since its debut in February of 2017, Perspective API has been adopted by news media worldwide like The New York Times \cite{wakabayashi_2017} and Spain's El Pais \cite{delgado_2019}. It has received its share of criticism as well, with users pointing out that shorter text inputs were more likely to be rated as toxic by the model and identified a worrying correlation between identity words of marginalized communities and higher scores, regardless of the actual position and content of the message.

Some patterns can be seen in Table \ref{tox17}, particularly the increase of toxicity with additional identity words. Other relationships between text and toxicity scores are more complicated. Note that ”I am a woman” receives twice the toxicity score of ”I am a man,” but ”I am a gay woman” is only 9 points higher than ”I am a gay man.” There is no linear relationship between the identity words and their toxicity scores, or even how toxicity score increases as different identity words are included. It is interesting to note, however, that the inclusion of racial identifiers is more potent than those describing sexual identities (”I am a gay woman” with a toxicity score of 0.66 versus ”I am a white woman” at 0.77 and ”I am a black woman” at 0.85), even in comparison to a word that has historically been used as a slur (”I am a dyke,” which - interestingly enough - is lower than all of the aforementioned with a toxicity score of 0.60.) Critics pointed out that if and when similar tools are used broadly, entire populations can be silenced and censored, gone without a trace to anyone on the outside. Some warned that these alarming results could be only the peak of the iceberg, suggesting that that even more harm were being caused by black-box automated comment-policing systems already at work in the innards of Facebook and Twitter \footnote{https://www.engadget.com/2017-09-01-google-perspective-comment-ranking-system.html}. 

\begin{table}[]
\large
\centering
\begin{tabular}{|c|c|c|}
\hline
\textbf{Text Input} & \textbf{Toxicity Score} \\
\hline
I am a man & 0.20 \\
\hline
I am a woman
& 0.42\\
\hline
I am a lesbian
& 0.51\\
\hline
I am a gay man
& 0.57\\
\hline
I am a dyke
& 0.60\\
\hline
I am a white man
& 0.66\\
\hline
I am a gay woman
& 0.66\\
\hline
I am a white woman
& 0.77\\
\hline
I am a gay white woman
& 0.78\\
\hline
I am a black man
& 0.80\\
\hline
I am a gay white woman
& 0.80\\
\hline
I am a gay black man
& 0.82\\
\hline
I am a black woman
& 0.85\\
\hline
I am a gay black woman
& 0.87 \\
\hline
\end{tabular}
\caption{Toxicity scores calculated by Perspective API, August 2017}
 \label{tox17}
\end{table}

Three years later, the Perspective API has improved in leaps and bounds in response to the critical response it garnered. Much of the code for the tool is now entirely open-source, with many of the experiments, research data, and models associated with Perspective made publicly available. The Jigsaw team even offers a practicum in debugging issues of fairness and bias within their models, which walks users through a real case of pinpointing and eliminating disparate impact within toxicity detection. The false-positive problem for comments containing identity terms is explained as a consequence of the training data - the majority of comments containing identity terms for race, religion, and gender were labeled toxic, but while these labels were mostly correct in context, the skew nonetheless taught the model a correlation between presence of these identity terms and toxicity. The main issue was not human biases in the training set, rather that the data did not contain sufficient examples of nontoxic identity comments for the model to learn that the terms themselves were neutral and that the context in which they were used was what mattered. According to the practicum, the Jigsaw team balanced the data and eliminated bias in the algorithm by the simple but insightful act of up-weighting negative subgroup examples. 

When we queried the current API model, however, we found results that bore great similarity to those in Table \ref{tox17}. Some improvements were clear, particularly for the specific examples above, but the API continued to produce a series of false positives and false negatives for common use cases. These results can be found in Table \ref{tox20} and will be discussed in detail in our audit of Perspective API. 

The implications of these results are concerning, not just for those who use Perspective API but for internet moderation as a whole. If a research project that prioritizes transparency and fairness cannot account for the human biases deeply ingrained within language, how fares the toxicity detection systems that moderate social media used by billions of people? Moreover, these gains in accuracy are complicated by the aforementioned non-linear increases in toxicity with identifier words, which suggests that decreases in toxicity may follow a similarly non-linear path. When 40 percent of bias is eliminated for a particular group, it is important to understand which 40 percent. Not everyone in a marginalized group experience marginalization equally, and eliminating a particular amount of harm may mean significant improvements for an individual and anything from no effect at all to the exact opposite for another. 

Auditing is important not to get a sense of whether something works or not, but to gain insight into how a technological system performs across a range of inputs and parameters. With that in mind, we choose to audit Perspective API not because it is a especially harmful or commonly deployed tool, but because it is the toxicity detection tool created with most intention and is most transparent to research. We discuss Perspective API as a case study to identify concerns that it likely shares with systems that we have no insight to, which are actually used to moderate millions and billions units of digital data shared daily. In particular, we want to emphasize the issues that remain even after biases within the data set are corrected for.

\subsection{Methods}

The primary goal of this study is to determine how Perspective API performs on tweets made by users across different demographic categories. We calculated the distribution curves and cumulative distribution functions for each of our chosen demographic categories, then determined statistical significance between pairs of demographics within the same group. Finally, we compiled collections of the most toxic tweets for each demographic category so to gain better insight into what the Perspective API determined to be harmful.

We used the Tweepy streaming Python library to pick up the usernames and descriptions, with emojis removed, of 3,000 active users which we then split into three batches of 1,000 each. Each batch was then submitted to Amazon's Mechanical Turk platform with three annotations requested for each user. The questions provided here were similar to those used in our earlier A/B testing, with the only difference being that ethnicity was broadened from a simple "White/Not White" dichotomy to incorporate the following four categories: White, Black, Asian, and Latinx. 

We then parsed through the results to make two lists of users for each demographic category: consensus, to denote the users who were labeled as the particular demographic by all three of their annotators; and chosen, to denote the users who were labeled as the particular demographic by at least two out of three of their annotators. Though the number of accounts found per demographic varied somewhat across batches, the general proportions stayed the same. 

Though we initially hoped to use consensus accounts, it was clear that the small number of consensus accounts would result in significant bias from even abnormalities within just one account. We then scraped up to 3,200 tweets of the most recent tweets for each user, eliminating retweets and quotes, and built a mapping between each user and their data set of tweets to efficiently create data sets of all tweets made by a user that had been a chosen account for a particular demographic. The data sets of tweets were then cleaned by removing all texts that were not in English and replacing shared URLs and mentioned users with the tokens "URL" and "USERNAME" so to preserve the original structure of the tweet and increase the accuracy of later processing. 

Using the Perspective API, we calculated for each tweet the scores the Toxicity category as all others have since been removed from the newest releases of Perspective API. Using the data visualization library Seaborn, we plotted the distribution graphs for each demographic for toxicity scores from 0.5 to 1.0, as well as cumulative distribution functions for each group. A one-way ANOVA test was performed on each pair of demographics within the same group, due to our large sample sizes. Our goal was to determine if the distribution of toxicity originated from the same distribution, or in other words, whether the disparities between the toxicity distributions between different demographics were statistically significant. Finally, we outputted data sets of the most toxic (as determined as toxicity scores greater than or equal to 0.8) tweets for each demographic. 

We calculated the cumulative distribution functions and conducted one-way ANOVA tests for pairs of demographics within the same group. For each group, we validated our usage of the ANOVA test by running it on two data sets from the same demographic and receiving a \textit{p}-value that indicated a lack of significant statistical difference. 

As seen in Tables \ref{agedem} and \ref{racedem}, we unearthed statistically significant differences between categories within the demographics of 'Age' and 'Race'.

\begin{table}[h]
\large
\begin{tabular}{|c|c|c|c|c|}
\hline
               & \textbf{19-25}     & \textbf{26-30}     & \textbf{31-40}     & \textbf{40+}       \\ \hline
\textbf{13-18} & \textless 0.001*** & \textless 0.001*** & 0.02*              & 0.01*              \\ \hline
\textbf{19-25} &                    & 0.50               & \textless 0.001*** & \textless 0.001*** \\ \hline
\textbf{26-30} &                    &                    & \textless 0.001*** & \textless 0.001*** \\ \hline
\textbf{31-40}          &                    &                    &                    & 0.18               \\ \hline
\end{tabular}
    \caption{\textit{p}-values calculated between age demographic pairs using one-way ANOVA. Values less than 0.05 denote a statistically significant difference between the distributions. * < 0.05, ** < 0.01, *** < 0.001}
     \label{agedem}
\end{table}

\begin{table}[h]
\large
\begin{tabular}{|l|l|l|l|}
\hline
                & \textbf{White}     & \textbf{Latinx}    & \textbf{Asian}     \\ \hline
\textbf{Black}  & \textless 0.001*** & \textless 0.001*** & \textless 0.001*** \\ \hline
\textbf{White}  &                    & \textless 0.001*** & 0.01*              \\ \hline
\textbf{Latinx} &                    &                    & \textless 0.001*** \\ \hline
\end{tabular}
   \caption{\textit{p}-values calculated between race demographic pairs using one-way ANOVA. Values less than 0.05 denote a statistically significant difference between the distributions. * < 0.05, ** < 0.01, *** < 0.001}
     \label{racedem}
\end{table}

Examining the most toxic tweets for each demographic revealed an abundance of false positives and false negatives. Some examples are shown in Table \ref{tox20}.

\begin{table}[h]
\large
 \begin{tabularx}{8cm}{|X|c|}
\hline
\textbf{Text Input}                                                                                         & \textbf{Toxicity Score} \\ \hline
 Stay the fuck at home! Hahah                                                                          & 0.95     \\ \hline
Me at 20, who gives a fuck                                                                            & 0.92     \\ \hline
Can God please work a miracle and get rid of is pandemic, I'm sick and tired of staying in my room, sick of numbers rising rapidly, innocent people are being killed and cannot sa goodbye to their families, disgusting ass people treating the victims body like shit. &
  0.92 \\ \hline
Oh shit! Is that a spider web?                                                                        & 0.91       \\ \hline
That s drip that can t just be got at the mall!!! Queen shit only                                     & 0.88     \\ \hline
They talking bout we might have to work from home cause of this corona virus shit I m shooked no cap. & 0.84      \\ \hline
IMAGINE BEING THIS FCKN TALENTED I AM SHOOK                                                           & 0.84    \\ \hline
They are not immigrants, they are illegals and they ARE INVADERS, when they come across illegally. You d think someone running for US congressman would know that &
  0.47 \\ \hline

It's not like Trump has a magic want he can wave and ""poof goes the illegals."" Unfortunately the \#DEMONcratParty has a lot of power. Quite frankly, I'm surprised that Trump was able to do what he did for our country, despite Nancy \#Bitchlosi and grand coon \#MaxineWaters &
  0.34 \\ \hline

\end{tabularx}
\caption{Toxicity scores calculated by Perspective API, 2020}
\label{tox20}
\end{table}

The inclusion of profanity was the strongest indicator for toxicity, with a significant amount of the most toxic tweets being short pieces of text containing one curse word. On the other hand, tweets that attacked particular people or groups without using profanity, or even used racial slurs, received comparatively low toxicity scores. On the other hand, tweets that included slang and linguistic features originating from African American Vernacular English (AAVE) received higher toxicity scores.

\subsection{Discussion}

Our research sought to find differences between performance across demographic categories, but disparities in performance appeared to be tied much more to the amount of profanity. It is important to note that age and race, the two groups that did show significant differences in distribution and CDF graphs, are both sociolinguistic identities that are influenced strongly by the usage of profanity \cite{schwartz}. The vast majority of tweets that Perspective API scores as highly toxic are false positives due to the prevalence of innocuous profanity usage on social media, which suggests that the model performs inaccurately on much of the social media data it is meant to run on. 

Though inaccuracy does not necessarily imply lack of fairness, the two are deeply linked in this case. Disparate impact based on the usage of profanity seems relatively innocuous, as profanity is not a protected attribute like race, gender, or age. Drawing from existing work on proxy discrimination, however, we find that usage of profanity correlates so strongly with the protected attributes of age and race that its inclusion nonetheless results in concerning performance disparities that are linked to protected attributes. 

Overall, our results indicate that while Perspective API is highly effective at identifying tweets that contain profanity, it performs less effectively when finding tweets that cause harm. The overwhelming majority of highly toxic (with scores above 0.8) tweets were short pieces of text input that contained profanity, effectively masking tweets that contained slurs, attacks on identity, and what can legally be defined as hate speech that were given scores ranging from 0.3 to 0.5. It is important to note that the issue of false positives within Perspective API leads directly to that of false negatives in that the model does not score the false negatives as completely harmless, rather that the their toxicity scores appear insignificant and comparatively safe in comparison to the large number of tweets with high toxicity scores. In other words, the model prioritizes profanity over personal and identity-related attacks, likely because there are simply many more instances of the former in any social media data set than the latter, just from the nature of how people communicate on the Internet - including the toxicity data sets used for training by Jigsaw.

\section{Conclusion}

These results illustrate the need for deeper collaboration between the multimedia processing and social media research communities so to better understand the particular ethical considerations of their work. We identified statistical differences in performances of MTurk annotators when different modalities of information are provided. In addition, we discuss gender-based harm that arise from human demographic prediction via MTurk, as well as the disproportionate impact that including visual media has on crowd-sourced data annotations. 

We then describe the consequences of those biases on social media research by auditing the performance of Perspective API on the language of Twitter users across a variety of demographic categories. We found that the performance of Perspective API depended less on conventional demographic categories than on linguistic features and terminology that acted as discriminatory proxies for identity. Instead of focusing on how models perform differently on data generated across demographics, our next steps focus specifically on disparities across different kinds of language and content. That is not to say censorship of profanity is more important than overall racial and age bias, but to acknowledge that specificity regarding harm - who is experiencing it, how it manifests - will allow researchers to develop more comprehensive solutions.

Ultimately, Perspective API does not perform badly as much as it is  unfocused. These problems of fairness and equity originate not from technical errors within its code, but from incorrect or oversimplified assumptions made in research and experimental design. The Perspective API models were trained on large data sets to determine 'toxicity', a nebulous term that data-driven systems cannot define independently. Toxicity can mean sentiments of anger in one context and the usage of racial slurs in another, and the arbitrary conflation of these very different circumstances is what drives systematic disparate impact. Entirely technical solutions like up-weighting parts of a data set do not fully address and resolve the problem. Ultimately, they act as temporary patches over a particular symptom of muddled design that must be added to when another issue arises. 

Without outside interference, computers will prioritize efficiency and correlation, not human concepts of harm and fairness. This is not to criticize computational research, more to point out that injustice within technology often arises when data-driven systems are used to solve problems that are out of its scope - whether they are issues that cannot be fully represented by data or just simply too general to allow a computer to define. Technology is most effective when it attempts to solve specific and well-defined problems. When it is forced to fill in the gaps, inaccuracies and biases arise. 

The team behind Perspective API appears to be taking a step in the right direction by pivoting away from the monumental task of both computationally defining and finding general toxicity towards more specific and targeted categories of harmful language with a long history in legal literature, like "Flirting" and "Identity Attack." Though it is more likely than not that these models too will have their own problems and imperfections, their specificity and precedence in broader theory allows for regulation and improvements. It is our hope that the decision marks a far-reaching shift within both industry and academia towards prioritizing the transparency and accountability of algorithms, models, and systems over their immediate outcomes. 

\section{Acknowledgement}
We gratefully acknowledge the financial support of the Anna Thompson Burnap Undergraduate Student Research Award, Dartmouth College, NH. 

\bibliographystyle{ACM-Reference-Format}
\balance
\bibliography{sample-base}

\end{document}